\documentclass{iopart}
\usepackage{iopams}
\usepackage{graphicx,color,times,bbm,psfrag,dsfont}
\usepackage[T1]{fontenc}
\usepackage[latin9]{inputenc}
\usepackage{makeidx}
\usepackage{subfigure}
\usepackage[english]{babel}
\usepackage{epsfig}

\newcommand{\bra}[1] {\langle{#1}\vert}
\newcommand{\ket}[1] {\vert{#1}\rangle}

\begin{document}

\title{Dirac Equation For Cold Atoms In Artificial Curved Spacetimes}

\author{O Boada,$^1$ A Celi,$^1$
JI Latorre,$^1$ and  M Lewenstein.$^{2,3}$}

\address{$^1$ Dept. d'Estructura i Constituents de la Mat\`eria,
Universitat de Barcelona, 647 Diagonal, 08028 Barcelona, Spain}
\address{$^2$ ICFO-Institut de Ci\`encies Fot\`oniques, Parc Mediterrani de la Tecnologia, E-08860
Castelldefels (Barcelona), Spain}
\address{$^3$ ICREA - Instituci\'o Catalana de Recerca i Estudis Avancats, 08010 Barcelona, Spain
}

\begin{abstract}
We argue that the Fermi-Hubbard Hamiltonian describing the physics
of ultracold atoms on optical lattices in the presence of
artificial non-Abelian gauge fields, is exactly equivalent to the
gauge theory Hamiltonian describing Dirac fermions in the
lattice. We show that it is possible to couple the Dirac fermions to an "artificial" gravitational field, i.e. to
consider the Dirac physics in a curved spacetime. We identify the
special class of spacetime metrics that admit a simple realization
in terms of a Fermi-Hubbard model subjected to an artificial
$SU(2)$ field, corresponding to position dependent hopping
matrices. As an example, we discuss in more detail the physics of the 2+1D
Rindler metric, its possible experimental realization and
detection.
\end{abstract}
\maketitle

\section{Introduction}

The studies of ultracold quantum matter in artificially designed
external gauge fields is one of the most rapidly developing areas
of physics of ultracold atoms
\cite{Fetter09,BZD08,Lewenstein07,Nathan-book,Dalibard10}. Originally, these
studies arose from investigations related to the response of superfluids, such
as Bose-Einstein condensates (BEC), to rotation (cf.
\cite{Pitaevskii03,Madison00}). On one hand, rotation induces
quantized vortices and/or vortex Abrikosov lattices
\cite{Abo-Shaeer01}. On the other, its effects are equivalent to
those of an artificial constant (Abelian) magnetic field. The latter
analogy has led to the idea of realizing strongly correlated quantum
liquids, such as the celebrated Laughlin liquid of the fractional
quantum  Hall effect (FQHE) (cf. \cite{Cooper01}) by means of
rapid rotation \cite{Wilkin00,Paredes03}. Unfortunately, reaching
the FQHE regime with rotation is experimentally very difficult; it
has been achieved recently,  but in a system of only $1<N\le 10$
 atoms in rotating microtraps (lattice
site potential wells) \cite{Sarajlic09,Gemelke10}. Several
researchers proposed, thus, alternative approaches involving for
instance laser-induced gauge fields that employ dark states in
3-level systems \cite{Juzeliunas04}, or laser induced gauge fields
acting on atoms confined in an optical lattice
\cite{Jaksch03,Mueller04,Sorensen05}. Other approaches concerned
rotating optical lattices \cite{Cornell06,Bhat06},  or
interactions of lattice atoms embedded in a rotating BEC
\cite{JK08}. Interestingly, the proposals involving lasers can be
relatively straightforwardly generalized to particles possessing
internal "color" states subjected to artificial non-Abelian gauge
fields \cite{Ruseckas05,Osterloh05}.

In the last two years there has been  a large number of works
reexamining these ideas and proposing experimental realization
within the reach of the current state of the art. The NIST group employed
an approach similar to \cite{Juzeliunas04} and used Raman (Bragg)
transitions in  Sodium to realize experimentally the first non-zero
constant vector (corresponding to zero "artificial" magnetic field) \cite{Y.-J.Lin09a}. Note that when the
gauge symmetry is provided by {\it Nature}  only gauge invariant
observables are physical. In the case of artificial gauge
potentials the situation is in principle different. Gauge
potential are controlled by the experimenters and the measurements of
quantities that depend on the choice of gauge 
are possible; in another words, the very process of the creation of
the gauge potential is not gauge invariant, even though the
resulting Hamiltonians are (for discussion see \cite{Boada2009,Boada2010} and \cite{Moller2010}).
NIST group was also able to generate vortices using the same
scheme with a potential configuration corresponding to a non zero
artificial magnetic field \cite{Y.-J.Lin09b}.  Several practical
extensions of the scheme of \cite{Juzeliunas04} were discussed in
  \cite{Spielman2009,Gunter2009}. New schemes were proposed
for alkali and earth-alkali atoms in optical lattices employing
superlattice techniques \cite{Gerbier09,Bermudez2010}, and on
atomic chips \cite{Goldman10}. Very recently, the creation of spin-orbit couplings, a special instance of synthetic
non-abelian fields, was reported \cite{SpielmanDarpa}.

Artificial non Abelian gauge fields are particularly interesting,
since they provide a natural framework to simulate relativistic
physics of the Dirac equation. Artificial Dirac physics has been in
recent years at the center of interest in condensed matter in
the context of studies of the amazing properties of graphene
\cite{Wallace47,Novoselov2005,McCann2006,Katsnelson2006,Zhou06,Geim2007,
Novoselov2007,Matulis2007,Jackiw2007,Pendry2007,Cheianov2007,CastroNeto2009}. In the case of graphene
the Dirac points in the dispersion relation appear due to the geometry
of the underlying hexagonal lattice. This idea can be carried over
to cold atoms \cite{Zhu2007}; a hexagonal optical lattice (OL) with spinor
bosons has been recently realized experimentally
\cite{Sengstock2010}. Also, other systems not relying on a lattice have been proposed to emulate Dirac particles \cite{unanyan10}. 
Other systems where Dirac physics plays a
role include superfluid 3He-A \cite{Volovikbook1,Volovikbook2} (where the analogy to Particle physics can be extended to include gauge interaction and Standard model phenomena, for instance see \cite{Voloviknature}), trapped ions \cite{Lamata2007,Bermudez2007,Gerritsma2009,Casanova2010,Gerritsma2010}, or Fermi-Bose mixtures
\cite{Cirac:2010us}.

In the case of artificial non-Abelian laser-induced fields (ANALF) the
connection to Dirac physics was  pointed out in  
\cite{Juzeliunas2008}. The Dirac equation is responsible for the anomalous integer
quantum Hall effect in artificial $SU(2)$ fields on a 2D square
lattice, and topological quantum phase transitions on 2D hexagonal
lattices, as well as FQHE \cite{Burrello:2010yz}.  These situations
correspond to relativistic physics in 2+1 dimensional spacetime
 \cite{Vaishnav2008,goldman:035301,Juzeliunas2010,Merkl2010}. Very recently, several
proposals were formulated for the creation of ANALF in 3+1 dimensional (3D spatial)
allowing for simulations of Wilson fermions \cite{Ginsparg1981} and
axion \cite{Wilczek78} quantum electrodynamics with ultracold atoms
\cite{Lepori:2010rq,Bermudez2010}.

So far, most of the proposals have dealt with constant non Abelian
fields strengths (i.e constant Wilson loops), staggered Abelian
gauge fields \cite{Lim2008,Lim2010}, or, in the rare cases, fields
that are linear in the spacial coordinates
\cite{Osterloh05,Satija06}. The Dirac equation, resulting in
some of these situations, corresponds to a Dirac equation in a
flat Minkowski space. The crucial ingredient of such studies was
based on the analysis of the dispersion relation between energy
and quasi-momentum, as in graphene \cite{CastroNeto2009}. Energy
bands touch in isolated singular points, called Dirac points. 
In the vicinity of these points, the dispersion relation linearizes,
and a cone is formed. Dirac physics can then be realized for
fermions adjusting the Fermi surface to include the Dirac point,
and considering {\it low lying} energy excitations.

In this paper we take a different perspective on this issue,
employing standard concepts from Hamiltonian lattice gauge theory (HLGT)
\cite{Montvay97}. We argue that using artificial non-Abelian
fields in lattices it is possible to simulate with cold atoms  a
Dirac spinor in the same way as it is done in HLGT. In another
words, the Hubbard Hamiltonian describing the physics of atoms in
artificial non-Abelian fields in lattices is exactly the HLGT
version of the Dirac's Hamiltonian.  That is to say, in the scheme
we consider, all excitations of the fermion field in the lattice
are Dirac-like, not only the low lying ones. The advantage of this
point of view is that it allows one to couple the simulated
Dirac-fields to external fields, or quantum fields in a
straightforward way, which is not so clear in artificial
graphene-like systems.

In particular, we will show that it is possible to consider coupling
Dirac fermions to an artificial gravitational field, that is
to consider the Dirac equation in curved space. We will identify
and focus on a special class of space-time metrics that admit a
simple formulation of the Dirac lattice Hamiltonian in terms of a
Fermi-Hubbard model subjected to an artificial $SU(2)$ field
\cite{goldman:035301}, corresponding to tunneling matrices with
position-dependent overall hopping rate.

We will not consider here  the fermion doubling problem as it is
inessential for the discussion of  gravitational effects
\footnote{The mixture of the two Dirac points due to the
gravitational background potentially induces a gauge field
coupling to the composite fermions, (for graphene like lattice see
for example \cite{Chamon2000,Jackiw2007,Pachos2007,Pachos2008}).
However, the field contribution is relevant in the presence of
conical singularities, disclination or dislocation in the graphene
language, while is negligible when the metric is smooth.}.
Feasible solutions of such a problem in OL simulations of 3+1
fermions have been very recently proposed in \cite{Lepori:2010rq}
and \cite{Bermudez2010}.

The motivation for simulating the Dirac equation in curved space-time
is at least two-fold. The propagation of fermions in curved space
is not {\it experimentally} accessible in high-energy
physics/cosmology as the (piece of the) Universe that we are nowadays able
to probe is practically flat \cite{wmap7}. The lattice
realization proposed in the paper is an appropriate description of
the continuum physics for processes occurring over scales of
several lattice spaces. The realization with cold atoms in OL
provides us with ``lap-top experiments'', paving the way to
observe exotic effects like the Thermalization Theorem, also known
as Fulling-Davies-Unruh effect \cite{Fulling73,Davies75,Unruh76}
(for a review on the subject see \cite{PTPS.88.1}). Roughly
speaking, it states that an accelerated observer perceives the
Minkowski vacuum as a thermal bath. This is a manifestation as
Hawking radiation \cite{Hawking74} of the same phenomenon: namely
the existence of a non-trivial Bogoliubov transformation between
the Minkowski vacuum and a space-time with a horizon. The effect
of the latter, in Quantum Information language, is to
``trace-out'' part of the system, giving raise to a thermal
reduced density matrix. Another exotic gravitational effect to
observe might be the  curved space-time version of Zitterbewegung
\cite{Singh2009}, to compare  with its recent observation achieved
in flat space \cite{Gerritsma2009}.

On other hand, simulating the Dirac equation in curved spacetime
gives rize to the possibility of modelling the analogues of graphene ripples
on a {\it flat, square} lattice. Such a possibility could help in
disentangling the action of ripples (which admit a natural
``geometric'' interpretations  \cite{Cortijo2007293} or a gauge field 
interpretations, e.g. \cite{Polini2010}) on the carrier
density from other contributions.

Different approaches to quantum simulation include the exact simulation
of the dynamics of a strongly correlated systems by unitary gates \cite{LCV2008}, or the 
the creation of interesting strongly correlated states which are ground states
of interesting Hamiltonians \cite{LPR2009}.

The paper is organized as follows. In section \ref{Dirac's
Hamiltonian on a Lattice} we demonstrate how the lattice Dirac
Hamiltonian in flat space-time \cite{Susskind77} can be obtained
as the discretization of the continuum Dirac Hamiltonian $H$. This
Hamiltonian is identified then as the Hamiltonian of the SU(2)
Fermi-Hubbard model considered  for instance in
\cite{goldman:035301}, where the hopping terms are given by the
Pauli matrices $\sigma_i$, $i= x,y$ times a constant hopping rate
$J$. Although the discretization can be done at this stage in many
different ways, we obtain it by coarse graining   the "symmetric"
formulation of $H$ in the fermion field $\psi$ and its conjugate
$\psi^\dagger$. This strategy turns out to be very convenient in
order to find the lattice version of the Dirac Hamiltonian in
curved space-time. As the first exercise, after reviewing  the
form of the continuum Hamiltonian on a generic manifold (admitting
a "time" isometry) in section \ref{Dirac's Equation in Curved
Spacetime}, we successfully apply the above strategy to the
interesting case of the 2+1 Rindler Universe \cite{Rindler66} in
section  \ref{An example: Dirac Hamiltonian in Rindler's
Universe}. The resulting lattice Hamiltonian differs from the
Hubbard Hamiltonian of the flat case in the following:  the
hopping rates exhibit a linear dependence with position. Such a
simple form is due to the cancellation of the spin-connection, once
$\psi$ and $\psi^\dagger$ are treated in the same manner: as a
consequence the hopping matrices need not be locally rotated.
We find that this simplification happens in fact for any static
spacetime. For the details of the derivation we refer the reader
to the  \ref{The generic case}. We discuss under which
conditions, and how the fermion propagation in such space-times
can be engineered and detected on optical lattices in sections
\ref{Dirac equation in curved spaces and optical lattices} and
\ref{Detection: The density of states at the Fermi level as simple
observable} that are the heart of the work. There, several
different experimental ways of implementing the Hubbard model of
interest are presented. We propose the density of states as the
relevant observable to capture the Dirac physics. It is a
measurable quantity both in graphene
\cite{Meyer2007,Meyer2007101,Stolyarova2007,Ishigami2007} and in
OL \cite{1367-2630-12-3-033041,GreinerBloch1,GreinerBloch2,GreinerBloch3,PolzikEckert1,PolzikEckert2,Oriol,Sengstock} experiments. We compute the theoretical value of
the density of states analytically in the continuum limit using
perturbation theory.

We collect our concluding remarks and discuss further developments in section \ref{Conclusions and Outlook}.

\section{Dirac's Hamiltonian on a Lattice}\label{Dirac's Hamiltonian on a Lattice}
In this section we show, by discretization the spacial coordinates, that Dirac's Hamiltonian in $2+1$ dimensions is a hopping
Hamiltonian with non-Abelian tunnelling matrices. To this end, let us recall that Dirac's equation in 2+1 dimensions reads $\gamma^a \partial_a \psi
= 0$, $a=0,1,2$, where $\gamma_{a}$ are Dirac's matrices
satisfying $\left\lbrace \gamma_{a},\gamma_{b} \right\rbrace
=2\eta_{ab} $, $\eta_{ab}$ is the metric, and $\psi$ is a two
component spinor. The time evolution of the field $\psi$ is,
\begin{equation}
i\partial_0 \psi = \underbrace{-i {\left( \gamma^0 \right) }^{-1}
\left(\gamma^1 \partial_1 +  \gamma^2 \partial_2\right)
}_{\mathcal{H}}\psi \, , \label{flatdirac}
\end{equation}
from which we can easily read-off the Hamiltonian. As was first pointed out
in   \cite{Susskind77}, in 2+1 dimensions and on a lattice of spacing $\Delta$ the discretized
version of equation (\ref{flatdirac}) is
\begin{equation}
\fl \qquad i\partial^0 \psi = -i\frac{ {\left( \gamma^0\right)
}^{-1}}{2\Delta}\left( \gamma^1 \left(\psi_{m+1,n} - \psi_{m-1,n}
\right) + \gamma^2 \left( \psi_{m,n+1} - \psi_{m,n-1}
\right)\right) \,,
\end{equation}
with $\psi({\bf x})=\psi(m \Delta,n \Delta)=\psi_{m,n}$. Rewriting
the Hamiltonians $H=\int dxdy \, \psi^{\dagger} \mathcal{H} \psi$
we obtain on the lattice,
\begin{equation}
\fl \qquad H=-i\frac{\left( \gamma^0\right) ^{-1}}{2\Delta} \sum_{m,n} \left( \psi^{\dagger}_{m+1,n} \gamma^1 \psi_{m,n} +\psi^{\dagger}_{m,n+1} \gamma^2 \psi_{m,n} \right) +h.c.
\label{dirac_flat_discrete}
\end{equation}
It is easy to see that (\ref{dirac_flat_discrete}) is nothing
more than a Fermi-Hubbard Hamiltonian with non-Abelian hopping
matrices where the interactions and the effect of the trap have
been neglected. In the notation of   \cite{goldman:035301} we
have $U_x = i{\gamma^0}^{-1}\gamma^1=i\sigma_1$ and $U_y =
i{\gamma^0}^{-1}\gamma^2=i\sigma_2$ (which implies
$\gamma_0=i\sigma_3$) and the Abelian flux $\Phi=0$.

For further reference, we note that the lattice Hamiltonian of (\ref{dirac_flat_discrete}) can be alternatively obtained by
computing
\begin{equation}
\fl \qquad  H=\int {{\rm d}} x{{\rm d}} y \, \psi^\dagger {\cal H} \psi =
 \frac 12\int {{\rm d}} x{{\rm d}} y \, ({\cal H} \psi)^\dagger \psi + \frac 12\int {{\rm d}} x{{\rm d}} y \, \psi^\dagger {\cal H} \psi\,,
\end{equation}
with the substitution  of spacial derivatives with finite
differences over one lattice space $\Delta$, $\partial_x \psi
\rightarrow \frac{\psi_{m+1,n}-\psi_{m,n}}{\Delta}$ and
$\partial_y \psi \rightarrow
\frac{\psi_{m,n+1}-\psi_{m,n}}{\Delta}$.

\section{Dirac's Equation in Curved Spacetime}\label{Dirac's Equation in Curved Spacetime}
In this section we review the formulation of Dirac's equation on a curved spacetime. Let $M$ be an 
arbitrary curved manifold and let us define a
\textit{vielbein} $e^{a}_{\mu}$, or set of vectors that form a basis of the tangent space $T_{M}$
at each point of $M$. Here the index $\mu$ labels the spacetime component and $a$ simply labels the basis vector.

Although $e^{a}_{\mu}$ is not constant in general, we require it to be covariantly constant (see for example \cite{Waldbook}).
We introduce a connection $\omega$ such that
\begin{equation}
\fl \qquad D_{[\mu} e^a_{\nu]} = \partial_{[\mu} e^a_{\nu]} + {\omega_{[\mu}}^a_b e^b_{\nu]} = 0\, .
\label{cov_drivein}
\end{equation}
This defines $\omega$ and the covariant derivative $D_{\mu}$. Dirac's equation for a Fermi field $\psi$ will be
\begin{equation}
\fl \qquad \gamma^{\mu} D_\mu \psi = 0\,,
\label{dirac}
\end{equation}
where $\gamma^{\mu}$ are the curved spacetime gamma matrices. The flat-space gamma matrices
$\gamma_{a}$ and the $\gamma^{\mu}$ are related by $\gamma_{\mu}(x) = e^{a}_{\mu}(x) \gamma_{a}$.
Thus,we have $\left\lbrace \gamma_{\mu}(x),\gamma_{\nu}(x)\right\rbrace  = 2 g_{\mu\nu}(x)$ if
$\left\lbrace \gamma_{a},\gamma_{b}\right\rbrace  =2 \eta_{ab}$ and the \textit{vielbein} forms
an orthonormal basis $ e^{a}_{\mu} e^{b\mu}=\eta^{ab}$, with $g_{\mu\nu}$ and $\eta_{ab}$ the
curved and flat-space metrics, respectively.

By separating the time component we have, $\gamma^{\mu} D_\mu = \gamma^t \partial_t  + \frac{1}{4}
\gamma^t \omega_t^{ab} \gamma_{ab} +\gamma^i \partial_i  + \frac{1}{4}\omega_i^{ab}\gamma_{ab}$. Now,
we are ready to write out the time variation of $\psi$,
\begin{equation}
\fl \qquad i\partial_{t}\psi = \underbrace{-i\left( \gamma^t \right) ^{-1} \left( \gamma^i \partial_i +\frac{1}{4}
\gamma^{i} \omega_i^{ab}\gamma_{ab}  + \frac{1}{4} \gamma^{t} \omega_t^{ab}\gamma_{ab} \right)}_{\mathcal{H}}  \psi \, ,
\label{curveddirac}
\end{equation}
where $\gamma_{ab} =\gamma_{a}\gamma_{b}$. 

Hence, we have identified the Hamiltonian for a Dirac fermion on a curved manifold. In section \ref{Dirac equation in curved spaces and optical lattices} we will proceed to its diagonalization analogously to the flat case and discuss its implementation on an optical lattice.

\section{An example: Dirac Hamiltonian in Rindler's Universe}\label{An example: Dirac Hamiltonian in Rindler's Universe}
As an example, we will derive Dirac's Hamiltonian in a particularly simple spacetime. Let us start by recalling that the Hamiltonian in curved spacetime,
 when a timelike killing vector is present, is the integral of the Hamiltonian density on a timelike hypersurface.
In 2+1 dimensions we have,

\begin{equation}
H=\int d\Sigma^2 \, \bar\psi \gamma^t\mathcal{H} \psi \, \label{H}.
\end{equation}
The differential is the volume element on a timelike slice and includes the determinant of the  metric, $d\Sigma^2 = \sqrt{-g} dx dy$.

In Rindler's space, the metric takes the form

\begin{equation}
ds^2 = -(ax)^2 dt^2 + dx^2 + dy^2 \,. \label{mrindler}
\end{equation}
We are interested in this metric for two reasons. First, it is the metric seen by an observer in constant acceleration in flat
spacetime. Therefore, it could have implications for earth-dwelling detectors observing cosmic background radiation. Second, it is the near-horizon metric of a Schwarzschild black hole.

The Rindler metric suggests the choice of \textit{dreivein} $e^0 = \vert ax \vert dt$, $e^1=dx$, $e^2=dy$. Using (\ref{cov_drivein}) we may compute the spin connection, whose
only non-vanishing component is $w_{t}^{01} = a\frac{x}{\vert x \vert}$. Dirac equation (\ref{curveddirac}), greatly simplifies in this spacetime,
\begin{equation}
i\partial_{t}\psi = \underbrace{-ia|x|\left(- \gamma_2\left( \partial_x +\frac{1}{2|x|}\right) + \gamma_1\partial_y  \right)}_{\mathcal{H}}  \psi \, . \label{hrin}
\end{equation}
where we have used that $\gamma_0 \gamma_1 \gamma_2 = -1$,  which holds only in 2+1 dimensions. In what follows, we adopt the gamma matrices representation choice $$\sigma_x=-\gamma_2,\ \ \  \sigma_y=\gamma_1,\ \ \  \sigma_z=-i\gamma_0.$$  In order to carry out the discretization analogously to how it is done for the case of a gauge theory, we note that the Hamiltonian can be written in terms of the Hamiltonian density (\ref{hrin}) as
\begin{eqnarray}
\fl \qquad H &=& \int {{\rm d}} x{{\rm d}} y \, \psi^\dagger {\cal H} \psi =  \frac 12\int {{\rm d}} x{{\rm d}} y \, ({\cal H} \psi)^\dagger \psi + \frac 12\int {{\rm d}} x{{\rm d}} y \, \psi^\dagger {\cal H} \psi  \cr
\fl \qquad &=& \frac{i a}{2} \int {{\rm d}} x{{\rm d}} y \, x   \left((\partial_x \psi^\dagger) \sigma_x \psi + (\partial_y \psi^\dagger) \sigma_y \psi - \psi^\dagger \sigma_x \partial_x\psi - \psi^\dagger \sigma_y \partial_y\psi\right).
 \label{Href}
\end{eqnarray}
In this symmetric form the spin-connection term disappears and it turns out again once integrating by parts.

The discretized version of the Hamiltonian (\ref{Href}) is simply obtained by the substitution $\partial_x \psi \rightarrow \frac{\psi_{m+1,n}-\psi_{m,n}}{\Delta}$ and $\partial_y \psi \rightarrow \frac{\psi_{m,n+1}-\psi_{m,n}}{\Delta}$, with $x=\Delta m$ and $y=\Delta n$. One readily gets
\begin{equation}
H = \frac{ia}{2}\sum_{m,n}m\left( \psi_{m+1,n}^\dagger\sigma_x \psi_{m,n} + \psi_{m,n+1}\sigma_y\psi_{m,n}\right)+H.c.\,.\label{Hrd}
\end{equation}
Therefore, a lattice with hopping matrices given by $J_iU_i = im\sigma_i$, $i=x,y$, growing linearly in the $x$ direction gives an appropriate description of free massless fermion in a Rindler spacetime. Such an Hamiltonian can be in principle implemented in a OL.
This problem will be tackled in the next section.

\section{Dirac equation in curved spaces and optical lattices}\label{Dirac equation in curved spaces and optical lattices}
 %{\bf XXX Insert a section with the resume of the material in the appendix and discuss which curved spaces the dirac equation can be coupled to in OL}

In view of the explicit realization in a OL we focus to the classes of spacetime where the massless fermion propagation can be described by the Hubbard model of the form
\begin{equation}
H=\frac i2 \sum J_{mn}\left(\psi^\dagger_{m+1,n}\sigma_x\psi_{m,n} +\psi^\dagger_{m,n+1}\sigma_y\psi_{m,n}\right) + h.c.\,.\label{curvedhubbard}
\end{equation}
As discussed in full details in the  \ref{The generic case}, we find that the lattice Hamiltonian reduces to this simple form only if the metric is static. In 2+1 dimensions it is equivalent to say that it can be written (in a certain coordinate system) in a diagonal form as
 \begin{equation}
 ds^2=- f^2(x,y) dt^2 + f^{-2}(x,y)e^{2\Phi(x,y)}(dx^2+dy^2)\,.\label{staticmetric}
\end{equation}
With this choice of coordinates the hopping rate is simply given by $J_{mn}=e^{\Phi(x_m,y_n)}$.

This is not the only requirement to be satisfied to reproduce such propagation on a OL, however.  Indeed, it is important to note that, even if the Dirac Hamiltonian written in the symmetric form (\ref{curvedhubbard}) is the same for any choice of the function $f$ in (\ref{staticmetric}), the corresponding Hamiltonian system is distinct for each metric as the canonical momentum is
$$\Pi=i\sqrt{-g} \bar \psi \gamma^t=i f^{-2} e^{2\Phi} \psi^\dagger,$$
and depends explicitly on $f$. Now, in optical lattice experiments the canonical momentum is fixed by the anticommutation relation to be simply $i \psi^\dagger$ that implies $f=e^\Phi$. Thus the cold fermions in the optical lattice simulate the propagation of massless fermions in a metric of the form \footnote{It is worth noting that for a generic function $\Phi$ the metric is curved and not Weyl-invariant, cfr \cite{I2010}.} 
\begin{equation}
ds^2= - e^{2\Phi} dt^2 + dx^2+dy^2\,. \label{timedef}
\end{equation}
The Rindler metric (\ref{mrindler}) corresponds to $\Phi= \log(ax)$.

%% I don't know if it is right place for these considerations

Before moving to the explicit implementation of the Hamiltonian
(\ref{curvedhubbard}) in OL, let us briefly discuss to what
extent it is a good description of the dynamics of massless
fermions in a space time given by (\ref{timedef}). There are two
kind of limitations. On one hand, due to finite size of the OL we
are able to cover only a finite portion of spacetime. On the other
end, the lattice approximation is valid when the metric, or the
function $\Phi$, is sufficiently smooth and slowly varying over
one lattice space $\Delta$. It is worth to note that the second
problem can be circumvented by using techniques from lattice gauge theory
(the continuum limit is obtained by extrapolation).

These two limitations should not obscure the physical content,
however, as OL with up to $300\times 300$ sites can be achieved.
This implies that the overall variation of the metric over lattice
can be of order one. Further considerations in the case of Rindler
space are given in section \ref{Conclusions and Outlook}.

\subsection{Experimental realization}

In this section we discuss briefly how one can achieve the
appropriate site-dependent hopping rate $J_{mn}$ in an OL. We
propose two different techniques.

%\subsubsection{a)Microwaves}
Recently it was proposed in   \cite{Mazza10,Bermudez2010},
that the intensity of the hopping can be tailored almost at will by
considering bichromatic spin-independent superlattices that trap
the hyperfine states of alkali bosonic or fermionic atoms. The
split of the hyperfine levels is controlled by a magnetic field.
The hopping between neighboring Zeeman sublevels of the $F$
(lower) hyperfine manifold, i.e. our "electrons", is induced via
adiabatic elimination of an intermediate $F=F\pm 1$ (upper)
hyperfine manifold coupled to $F$ via an off-resonant Raman
transition. For the details of the scheme for fermionic $^{40}$K
atoms, we refer to the original proposal \cite{Bermudez2010}.
$^{40}$K have $F=$9/2, and allow in principle to
simulate any pseudo-spin $F'\le $9/2, employing the
splitting of the Zeeman sub-levels in a magnetic filed, and optical
pumping to the relevant sub-levels. For the purposes of the
present paper it sufficient to  have $F'=1/2$, or alternatively to
use from the very beginning atoms with $F=1/2$ in the ground state
manifold, such as $^{6}$Li.

To be more concrete, consider a $^{6}$Li Fermi gas loaded on a 2D
square optical lattice of size $L\times L$, where the relevant
information of our quantum simulator is encoded in the Zeeman sub-levels of
the hyperfine manifold $F=$1/2. Laser-assisted
tunneling methods allow us to design  arbitrary operators
$U_{\boldsymbol{r} \boldsymbol{\nu}}$ dressing the hopping between
lattice sites $\boldsymbol{r}\to\boldsymbol{r}+\boldsymbol{\nu}$,
where
$\boldsymbol{r}=m\hat{\boldsymbol{x}}+n\hat{\boldsymbol{y}}$,
$m,n,\in\{1...L\}$,  and
$\boldsymbol{\nu}\in\{\hat{\boldsymbol{x}},\hat{\boldsymbol{y}}\}$.
Usually, such schemes  rely on Raman couplings to auxiliary states
trapped in the links of the original lattice and  belonging to a
different hyperfine manifold. Here, following
\cite{Mazza10,Bermudez2010} we use bichromatic spin-independent
superlattices , and use the secondary minima of
$F=$3/2 as bus states to mediate the hopping. Note
that the individual addressing of each hopping rate is granted by
the Zeeman splitting within the hyperfine manifolds, and the
different detunings of the Raman lasers. These  detunings can be
quite large, so that the  lifetimes of atoms on the lattice
(limited by photon absorption and spontaneous emission) can be
quite large, of order $\tau_{\mbox{l}}\sim 1$s. By making the Raman
laser intensity/detunings and/or Zeeman splitting spatially
dependent one obtains the desired spatially dependent hopping
rates which is necessary for the realization of curved space-times, equation
(\ref{curvedhubbard}). On top of that, is is possible to use
Feschbach resonances to turn off the atom-atom scattering, and
make the system essentially non-interacting.

%\subsubsection{b)Inverse engineering: simulating a curved space with laser waist}
An alternative, and perhaps even a simpler method to realize the
Fermi-Hubbard model of the form of (\ref{curvedhubbard}) can be
achieved by taking into account the finite waist of the lasers used
for the generation of the hopping terms. In general, this is an
undesirable feature,  and it can usually be neglected.  Indeed,
typically  Gaussian laser beams of waist $w$ are used,
characterized by a Rayleigh length ---the
the distance along the direction of propagation from the
waist to the place where the area of the cross section is doubled--- $z_R=\pi w^2/\lambda$
with $\lambda$ denoting the wavelenght.  That is to say, within a volume of $w^2\times z_r$, the  ideal planar wave is a
good approximation, at least around the center of the beam. For a
lattice with $L=30$, its linear size is $30\times \lambda/2$, so
that focusing a laser on the whole lattice $w\simeq 15\lambda$
leads to $z_R\simeq 700 \lambda$, so that the plane wave
description can be even used for an array of a few hundreds of 2D
$30\times 30$ lattices.

There are, however, no technical obstacles to focus the lasers on
much smaller spots, smaller than $L\lambda/2$, keeping $z_r$ still
quite large. Actually, such spacial modulation of the intensity
due to the waist of "real" lasers can be used to induce hopping
terms that depend non-trivially on the position \cite{Gunter2009}.
In general, the hopping rate, i.e. the modulus of the hopping
term, is proportional to the intensity of the laser producing it. For
instance, taking the paradigmatic example of the hopping induced by
the Raman transition in Jaksch and Zoller's setup \cite{Jaksch03}, an optical
lattice implementing the Hamiltonian  of (\ref{curvedhubbard})
can be achieved for Raman lasers propagating
 {\it all} in the same direction,
 once we consider only the radial waist and neglect the waist along the beam
\footnote{At the moment, it is not clear to us whether it is possible to engineer the
hopping matrices $\sigma_x$ and $\sigma_y$ with parallel and anti-parallel Raman laser.
Actually, we do not know of any explicit realization of such hopping in a setup \'a la Jaksch and Zoller.
Maybe, the dark state or slow-light method \cite{Juzeliunas04} is more promising.
The main point that the hopping rate is controlled by the intensity, remains valid.}. In this case, the shape
 of the laser intensity will correspond to the $e^\Phi$ factor of the metric.

% comparison between the 2:
Comparing the two methods explained above, the former has the advantage that, in principle, any shape of
 $J_{mn}$, i.e. any metric of the form (\ref{timedef}), can be engineered, but at the price of dealing with a quite
 involved experimental apparatus, while the latter, although it allows for a restricted choice of
 $J_{mn}$ (for instance a  Gaussian shape) is almost for free. The desired hopping rate profile is obtained by reverse engineering 
 of laser waist.

\section{Density of states at the Fermi level as simple observable}
\label{Detection: The density of states at the Fermi level as simple observable}

%{\bf XXX Is it better to make two sections for the theoretical derivation and the way of measuring the density? XXX}

Let us turn to the discussion of possible detection schemes  of Dirac
physics in curved spacetime. A simple observable  that
characterizes such physics, and that contain information about
the effects of the beam waist is the density of states
\cite{Cortijo2007293,Cortijo2009659,PhysRevB.76.165409}. This
quantity is routinely measured in graphene using scanning tunnel
\cite{Stolyarova2007,Ishigami2007} and electron
 transmission spectroscopy \cite{Meyer2007,Meyer2007101}. %{\bf In an optical lattice?}
For ideal graphene at zero temperature the density of states is
zero at the Fermi level and is proportional to $|E|$ (once we have
fixed $E_F=0$) as the charge carries are described by massless
fermions propagating in flat space. In the presence of deformations of
the graphene sheet its deviation from the free behavior can be
analytically computed at first order from the propagator of the
Dirac equation. Following \cite{Cortijo2007293},  by modeling such
deformations as perturbations of the Minkowski metric it is possible
to compute the Green function treating the correction to
 the free equation as an interacting term $V$. In fact, in this section we will first
  reproduce the computation of
 \cite{Cortijo2007293} for a metric of the form  (\ref{timedef}), instead of the spatial deformation
 (as in (\ref{spam})) considered there.

Our final goal is to determine the local density of states, defined by
\begin{equation}
 \int {\rm d} w \sqrt{-g}{\rm d}^2 r \rho(w,{\bf r})\equiv  \mbox{ \# of states }\,,
\end{equation}
 in terms of the Feynman propagator (see appendix) using the relation
\begin{equation}
 \rho(w,{\bf r})=\mbox{sign}(w)\frac 1\pi\mbox{Im}\left[\Tr \hat S_F(w,{\bf r},{\bf r})\gamma^t\right]\,, \label{localdensity}
\end{equation}
where ${\bf r}= (x,y)$.

In order to find $\hat S_F(w,\bf r,\bf r)$ we start by using the defining equation for the fermion propagator
\begin{equation}
 -i\sqrt{-g} \gamma^\mu \frac{D}{Dx^\mu} S_F({\bf x}, {\bf x'}) = \delta^3({\bf x} - {\bf x'}),
 \end{equation}
where the $\bf x$ indicates a point of the spacetime, to a metric
of the form (\ref{timedef}). By retaining
terms linear in $\Phi$, the above equation can be written as
\begin{equation}
-(i(\gamma^\mu \frac{\partial}{\partial x^\mu})_{Flat} + V({\bf x})) S_F({\bf x}, {\bf x'}) =\delta^3({\bf x} - {\bf x'})\,,\label{curvedpro}
\end{equation}
where
\begin{equation}
 V= i \gamma_1 \left(\Phi \partial_x + \frac 12 \partial_x \Phi\right) + i \gamma_2 \left(\Phi \partial_y + \frac 12 \partial_y \Phi\right)\,,
\end{equation}
is the effective ``external potential''. As we are interested to
$\hat S_F(w,\bf r,\bf r')$ and due to the time translation
invariance
$$S_F({\bf x}, {\bf x'})=S_F(t-t',{\bf r},{\bf r'})=\hspace{-.5mm}\int \frac{{{\rm d}} w}{2\pi} e^{-iw(t-t')}\hat S_F(w,\bf r,{\bf r'}), $$
 it is convenient to perform the Fourier transformation in time of  (\ref{curvedpro})
\begin{equation}
 \left(w\gamma_0 - i \nabla_{\bf r}\cdot \gamma -V({\bf r})\right)\hat S_F(w,{\bf r},{\bf r'}) =\delta^2( {\bf r}-{\bf r'})\,.
\end{equation}

The above equation can be solved consistently within the first order approximation by
\begin{equation}
 \hat S_F^1(w,{\bf r},{\bf r'}) =\int {{\rm d}}^2 r''  \hat S_F^0(w,{\bf r},{\bf r''}) V({\bf r''})\hat S_F^0(w,{\bf r''},{\bf r'})\,,\label{pro1}
\end{equation}
where
\begin{equation}
 \hat S_F^0(w,{\bf r},{\bf r'})= \int \frac{{{\rm d}}^2 k}{(2\pi)^2}\frac{w \gamma_0 - {\bf k \cdot \gamma}}{w^2-{\bf k}^2 + i \epsilon} e^{i \bf k (\bf r - {\bf r'})}\,,
\end{equation}
is the free fermion propagator and space translation's invariance
holds. By using the Fourier transformation of $\Phi({\bf r''})$
\begin{equation}
 \Phi({\bf r''})=\int \frac{{{\rm d}}^2 p}{(2\pi)^2} e^{i \bf p \cdot {\bf r''}} \Phi({\bf p})\,,
\end{equation}
equation (\ref{pro1}) can be explicitly computed performing the integration in ${\bf r''}$.

The relevant contribution to the trace turns out to be linear in $w$. Explicitly
\begin{equation}
 \Tr[\hat S_F^1(w, {\bf r} , {\bf r} ) \gamma_0]= \int \frac{{{\rm d}}^2 p}{(2\pi)^2} e^{i {\bf p} \cdot {\bf r} } \Phi( {\bf p} )  \Gamma(w, {\bf p} )\,,
\end{equation}
with
\begin{equation}
  \Gamma(w, {\bf p})=\int \frac{{\rm d}^2 k}{(2\pi)^2}  \frac{4w |{\bf k}-\frac 12 {\bf p}|^2} {(w^2-{\bf k}^2 + i\epsilon)(w^2 -({\bf k} - {\bf p} )^2 + i\epsilon)}]\,.\label{Gamma}
\end{equation}
The above integral is logarithmically divergent, but its imaginary
part is not. It is easy to show that this is the only part
contributing to the density. Indeed, as $\Phi({\bf p})$ is the
Fourier transformation of a real function, $\Phi({\bf p})^*=
\Phi(-{\bf p})$, and  $\Gamma(w, {\bf p})$ is even in ${\bf p}$,
$\Gamma(w, {\bf p})=\Gamma(w, -{\bf p})$, one finds
\begin{eqnarray}
 \left(\Tr[\hat S_F^1(w, {\bf r} , {\bf r} ) \gamma_0]\right)^*&=&\int \frac{{{\rm d}}^2 p}{(2\pi)^2} e^{-i {\bf p} \cdot {\bf r} } \Phi( {\bf p} )^*  \Gamma(w, {\bf p} )^*\cr
&=&\int \frac{{{\rm d}}^2 p}{(2\pi)^2} e^{i {\bf p} \cdot {\bf r} } \Phi( {\bf p} )  \Gamma(w, {\bf p} )^*,		
\end{eqnarray}
which immediately implies
\begin{equation}
 \delta \rho(w)=\mbox{sign}(w)\frac 1{\pi}\int {\rm d}^2 r \int \frac{{{\rm d}}^2 p}{(2\pi)^2} e^{i {\bf p} \cdot {\bf r} } \Phi( {\bf p} ) \mbox{ Im}\, \Gamma(w, {\bf p} )\,.
\end{equation}
The explicit  expression for $\mbox{ Im}\, \Gamma(w, {\bf p} )$ is
\begin{equation}
 \mbox{ Im}\, \Gamma(w, {\bf p} ) = 2 w -\frac w\pi \mbox{ Im}\left[\frac {\arctan\chi(w,p)}{\chi(w,p)}\right]\,,
\end{equation}
where $\chi(w,p)=\frac{p}{\sqrt{|4 w^2 -p^2|}\left(\Theta(4w^2-p^2) - i \Theta(-4w^2+p^2)\right)}$, and $p\equiv|{\bf p}|$. More details are given in the  \ref{The computation of Gamma}.

Hence, the density of states always receives a correction
proportional to $\Phi({\bf r})$ itself
\begin{equation}
\fl  \delta \rho(w)=\int {\rm d}^2 r \frac {2|w|}{\pi} \Phi({\bf r}) -\frac{|w|}{\pi^2}\int {\rm d}^2 r\int \frac{{{\rm d}}^2 p}{(2\pi)^2} e^{i {\bf p} \cdot {\bf r} } \Phi( {\bf p} ) \mbox{ Im}\,\left[\frac {\arctan\chi(w,p)}{\chi(w,p)}\right]\,.\label{deltarho}
\end{equation}

\subsection{Example: Density of states for a Gaussian beam}

We are now able to compute the correction to the density of state
when the hopping $J$ has a Gaussian shape due to the finite laser
beam waist. Under the assumption that the Raman lasers propagate
along $y$ direction, this implies that $\Phi(x,y)= -(\frac xa)^2$,
where $a$ is of order $10^2$ lattice spacings \cite{BZD08}. As
$\Phi({\bf p})= + \frac {(2\pi)^2}{a^2}\delta''(p_x) \delta(p_y)$,
the second term of (\ref{deltarho}) turns out to be zero and
the correction to the density of state simply reduces to
\begin{equation}
 \delta \rho(w)=\int {\rm d}^2 r \frac {2|w|}{\pi} \Phi({\bf r})\, \label{deltarhogau}
\end{equation}
providing a clear experimental signature. The local density of states gets a quadratic correction in $x$.

Incidentally, the same cancellation happens in case of an
exponential behavior of $J$, i.e. for a linear $\Phi({\bf r})$.
Consequently, the simple relation (\ref{deltarhogau}) applies
also to this case.

%{\bf XXX I can compute explicitly the correction to the density in the case of Rindler or any other but I don't know whether we are able to implement them XXX}

\subsection{Experimental detection}

A recent review of the detection methods that can be applied to
investigate Dirac physics with ultracold fermions in non-Abelian
gauge fields is contained an article authored by one of us
\cite{1367-2630-12-3-033041}. Here we just summarize this
discussion with particular focus on density of states. Let us
start by observing that for a non-interacting Fermi gas at $T=0$,
the total number of fermions $N_F=\int^\mu dE \rho(E)$, where
$\mu$ is the chemical potential, equal at $T=0$ to the Fermi
energy $E_F$. We see that $\rho(E_F)=dN_F/dE_F$ so that measuring
of the variance of $N_F$ with $E_F$  allows to determine
$\rho(E_F)$. If the systems is confined additionally in a slowly
varying harmonic potential $V({\bf r})$, a local chemical potential
can be introduced $\mu({\bf r})=E_F-V({\bf r})$, and the
corresponding local density of states, related to the  local
density by
$$n({\bf r})=\int dE n(E)\Theta(\mu({\bf r})-E),$$
where $\Theta(.)$ is the Heaviside (step) function. In this case
we get $$dn({\bf r})/d\mu({\bf r})=\rho(\mu({\bf r})),$$ i.e. a
similar formula to the Streda formula used in  
\cite{1367-2630-12-3-033041} for detection of Hall conductivity.
The determination of density of states can be achieved by:
\begin{itemize}
\item Measurements of the total number of fermions as a function of the chemical potential.
Here, the best currently available methods are: direct {\it in
situ} individual atom detection \cite{GreinerBloch1,GreinerBloch2,GreinerBloch3}, or quantum
spin polarization spectroscopy \cite{PolzikEckert1,PolzikEckert2}

\item Measurements of the (coarse-grained) local density  of fermions as a function of
the local chemical potential. Again, the best currently available
methods are: direct {\it in situ} individual atom
detection \cite{GreinerBloch1,GreinerBloch2,GreinerBloch3}, or quantum spin polarization
spectroscopy with spatial resolution \cite{Oriol}.

\item Measurements of frequency-momentum resolved single particle
excitation spectrum, such as those being done in Bragg (Raman)
scattering spectroscopy (for a state-of-the-art report see
\cite{Sengstock}). The spectrum in such processes is proportional to
the density of initial states of the scattering process.

\end{itemize}
Of course, many other methods, such as atomic ARPES, noise
interferometry, or even absorbtion and/or phase contrast imaging
can give at least indirect information about $\rho(E)$. All
of these methods are well developed in experiment with ultracold
atoms (see \cite{1367-2630-12-3-033041} and references therein).

\section{Conclusions and Outlook}\label{Conclusions and Outlook}

In this paper, we discussed the simulation of the Dirac equation
in artificial curved space-time with cold atoms. We showed that
using state-of-the-art techniques it is possible to simulate
relativistic fermion dynamics in curved spacetimes with a {\it
flat 2D square} lattice for an interesting class of 2+1 metrics.
Moreover, we pointed out the relation between a certain class of
Hubbard models and  Dirac's Hamiltonian in curved backgrounds, which
can be employed to make analytic computations in the continuum
limit of the former. We proposed to characterize the {\it Nature} of
Dirac fermions  on the lattice by measuring the density of states
at the Fermi level. This observable can be, on one hand,
analytically computed in perturbation theory in terms of Dirac
propagator, and, on the other hand, is accessible to measurements.

The present study opens the way to the direct observation of
elusive effects such as Rindler noise. Because we deal with odd
dimensional (2+1) Rindler system, the Dirac thermal noise,
measured by an ideal point-like De-Witt detector as a consequence
of the local acceleration, is expected to be ``anomalous'' (see
Ch.8 of \cite{PTPS.88.1}), i.e. it should follow Bose-Einstein
distribution. This issue is currently under investigation.

\section*{Acknowledgements}

A.C. thanks J.M. Pons and J. Garriga for useful discussions and J.G. Russo for giving insights into the thermalization theorem and for 
pointing out   \cite{PTPS.88.1}. M.L. acknowledges  MICINN (FIS2008-00784
and Consolider Ingenio 2010 QOIT), EU (AQUTE,NAMEQUAM), ERC
(QUAGATUA) and Alexander von Humboldt Foundation. O.B., A.C. and J.I.L 
acknowledge financial support from QAP (EU), MICINN (Spain), Grup consolidat (Generalitat de Catalunya), 
and QOIT Consolider-Ingenio 2010 is acknowledged. O.B. was supported by FPI grant number BES-2008-004782. 

\appendix

\section{Dirac Hamiltonian in spatially (graphene like) deformed metric}

Now we consider a different situation where the 2+1 metric is spatially non-trivial. Such a case is relevant in describing the properties of a graphene sheet with ripples. The most generic spacial deformation (at least in some patch) can be always written as
\begin{equation}
 ds^2= - dt^2 + e^{2\Phi(x,y)}(dx^2+dy^2).\label{spam}
\end{equation}

The {\it driebein} are
$$ e^0 =dt\, \ \ \  e^1= e^{\Phi(x,y)} dx\, \ \ \  e^2= e^{\Phi(x,y)} dy\, \ \ \  i=x,y\,,$$
and the spin-connection can be chosen to be non-trivial in the spacial part only
\begin{equation}
 w^{12}= \partial_y \Phi dx - \partial_x \Phi dy.
\end{equation}
It follows that the curvature is
\begin{equation}
 \Omega= 2 e^{-2 \Phi}\Omega^{12}_{xy} = -2 e^{-2 \Phi}(\partial_x^2 + \partial_y^2)\Phi\,.
\end{equation}
For instance, the slices of the metric (\ref{spam}) at constant time will be spheres or hyperboloids for $\Phi$ a positive or a negative quadratic form of $x$ and $y$, respectively.

Applying (\ref{curveddirac}) to this case we find
\begin{equation}
\fl \qquad  i \partial_i \psi={\cal H} \psi = -i\gamma_0e^{-\Phi}\left(\gamma_1(\partial_x + \frac 12 \partial_y\Phi \gamma_{12}) + \gamma_2 (\partial_y -\frac 12 \partial_x \Phi \gamma_{12})\right) \psi\,.\label{spcdh}
\end{equation}
We are tempted to interpret the above Hamiltonian as that of fermions coupled to a ``geometric'' non-Abelian  vector potential ${\bf A} \equiv (\partial_y \Phi,-\partial_x \Phi) \sigma_z$. Adopting the the gamma matrices' representation of the previous section, we can write it as
\begin{equation}
{\cal H} \psi = -ie^{-\Phi}\left(\sigma_x(\partial_x + i A_x) + \sigma_y(\partial_y + i A_y)\right)\psi\,,
\end{equation}
where the presence of $\bf A$ indicates that the rotation in the $xy$-plane, i.e. the $SO(2)$ subgroup of the Lorentz group, is promoted to a local symmetry in the background described by the metric (\ref{spam}). Such identification  is related to the treatment of the conical defects in the graphene sheets (dislocations and disclinations) as sources of magnetic fluxes. However, that this interpretation here is misleading as the gauge group does not commute with the spacetime symmetry, as $\sigma_z$ anticommutes with $\sigma_x$ and $\sigma_y$.

Taking in account this fact, the Hamiltonian density can more appropriately be written as
\begin{equation}
 {\cal H}_{(s.d.)} = -i\left(\sigma_x(\partial_x + \frac 12\partial_x\Phi) + \sigma_y(\partial_y + \frac 12 \partial_y\Phi)\right)\psi\,,
\end{equation}
where the symmetric role of $x$ and $y$ is evident.

Now we are ready to compute the total Hamiltonian. By rewritting (\ref{H}) we get
\begin{equation}
 H=\int {{\rm d}} x {{\rm d}} y\, e^{2\Phi(x,y)}\, \psi^\dagger {\cal H}_{(s.d.)} \psi\,.
\end{equation}
Using the the same manipulations as in the previous section we can recast it into a form where the spin-connection is not present,
\begin{equation}
 H=\frac i2 \int {{\rm d}} x {{\rm d}} y\, e^{\Phi(x,y)}\, \sum_{i=x,y}\left((\partial_i\psi^\dagger) \sigma_i \psi - \psi^\dagger\sigma_i \partial_i\psi\right) \,.
\end{equation}
 It follows that the discretized version of $H$, as in Rindler spacetime, takes the form of a SU(2) Fermi-Hubbard model with the modulus of hopping depending on the position
\begin{equation}
\fl \qquad  H_{FH}= \frac i2\sum_{m,n} \frac{e^{\Phi(m\Delta,n\Delta)}}{\Delta}\left( \psi_{m+1,n}^\dagger\sigma_x \psi_{m,n} + \psi_{m,n+1}\sigma_y\psi_{m,n}\right)+h.c\,.\label{HFH}
\end{equation}
At first sight it seems very surprising that the discretized Hamiltonian of massless fermions in Rindler geometry coincides with the one of fermions propagating in metric of the form of (\ref{spam}). Indeed, by taking $\Phi=\ln (a x) $ the expression (\ref{HFH}) reduces to  (\ref{Hrd}). Such an apparent contradiction  disappears upon closer inspection. At the end, from the point of view of the Dirac Hamiltonian for both metrics, what has changed with respect to the flat case is the effective speed of light, or equivalently the hopping rate, which becomes position (and direction) dependent.

Although the origin of the position-dependent hopping rate is different in the two cases, it comes from the Hamiltonian density in the Rindler case while is due to the invariant measure in the other, the effect is the same.
Roughly speaking, there are two possible way of modifying the effective speed of light in one spacetime direction, let us say $x$: one is to change $g_{tt}$ while the other is to change $g_{xx}$ by an inverse factor.

Nevertheless, the eigenfunctions and the spectra of the two Schr\"odinger problems remain different as the Hamiltonian densities in the two cases are.

\section{The generic case}\label{The generic case}

In order to treat the generic case let us retrace a few steps and analyze the formal expression for the Hamiltonian (\ref{H}).
First of all, we show that it corresponds to the Legendre transformation of the relativistic Lagrangian:
\begin{equation}
 L=\int {\cal L} = i\int \sqrt{-g} \,\bar \psi  \gamma^\mu D_\mu \psi\,,
\end{equation}
once we have chosen the coordinates to have that the timelike Killing vector is $K=\partial_t$, which implies $\partial_t g_{\mu\nu}=0$.
Indeed, by defining ${\cal{H}}=\frac{\delta {\cal L}}{\delta \partial_t  \psi} \partial_t \psi - {\cal L}$, the expression (\ref{H}) is recovered
\begin{equation}
\fl \qquad  H\equiv \int h = -i \int \sqrt{-g} \,\bar \psi \left( \gamma^i \partial_i + \frac 14 \gamma^\mu w_\mu^{\phantom \mu ab} \gamma_{ab}\right) \psi\,,\qquad i=x,y.
\end{equation}

Now, it is instructive to check explicitly that the Hamiltonian above is a Hermitian operator due to the existence of a timelike isometry. By using that $(\gamma^\mu)^\dagger =\gamma_0 \gamma^\mu \gamma_0$ and noticing that
$$\left(\bar \psi \,\gamma^\mu w_\mu^{\phantom \mu ab} \gamma_{ab}\, \psi\right)^\dagger = - \bar \psi \,w_\mu^{\phantom \mu ab} \gamma_{ab} \, \gamma^\mu\psi\,,$$
we get that
\begin{equation}
 H^\dagger = i\int \sqrt{-g} \left( (\partial_i \bar \psi) \gamma^i - \frac 14 \bar \psi\, w_\mu^{\phantom \mu ab} \gamma_{ab}  \gamma^\mu\right)\psi\,.\label{Hdag}
\end{equation}
In order to compare the above expression with $H$ it is convenient to integrate by parts and rewrite it as
\begin{eqnarray}
\fl \qquad H^\dagger &=& -i\int \sqrt{-g}\,\bar \psi \left(  \gamma^i \partial_i  + \frac 14 w_\mu^{\phantom \mu ab} \gamma_{ab}  \gamma^\mu  + \gamma^i\partial_i\ln \sqrt{-g} + \partial_i\gamma^i \right)\psi\cr
\fl \qquad &=& H  - i\int \sqrt{-g}\,\bar \psi \left(\frac 14 w_\mu^{\phantom \mu ab} [\gamma_{ab},\gamma^\mu] + \gamma^i\partial_i\ln \sqrt{-g} + \partial_i\gamma^i \right)\psi\,.
\end{eqnarray}

$H^\dagger$ and $H$ are the same if and only if the metric, and consequently the driebein, are invariant under time translation. Indeed, due to the conventional constraint $\nabla_\mu e_\nu^a +w_{\mu \phantom a}^{\phantom \mu a} e_\nu^b =0 $, it is true that 
\begin{equation}
e_\nu^a \nabla_\mu e^\rho_a = - e^\rho_a \nabla_\mu e_\nu^a= - e_{\nu a} w_{\mu \phantom a}^{\phantom \mu a} e^{\rho b}\,,
\end{equation}

which, using the commutator $[\gamma_{ab},\gamma^\mu]= 4 e^{\mu c} \gamma_{[a}\eta_{b] c}$, implies
\begin{equation}
 \frac 14 w_\mu^{\phantom \mu ab} [\gamma_{ab},\gamma^\mu] = \gamma_a w_\mu^{\phantom \mu ab} e_b^\mu = -(\nabla_\mu e^\mu_c) \gamma^c= -\nabla_\mu \gamma^\mu\,.
\end{equation}
  The cancellation follows from the identity
\begin{equation}
\partial_i\gamma^i + \gamma^i\partial_i\ln \sqrt{-g} =\nabla_i \gamma^i = \nabla_\mu \gamma^\mu\,,
\end{equation}
as $\partial_\mu\ln \sqrt{-g}= \Gamma_{\mu\nu}^\nu$ and $\partial_t\gamma^t = \gamma^t\partial_t\ln \sqrt{-g}=0$ if and only if $\partial_t g_{\mu\nu}=0$.

At this point, we can use expression (\ref{Hdag}) for $H^\dagger$  to get a Hamiltonian symmetrical in $\psi$ and $\psi^\dagger$.
By writing $H\equiv \frac 12 (H+H^\dagger)$ we find
\begin{equation}
 H=\frac i2 \int \sqrt{-g}\left(
\left(\partial_i \bar \psi \gamma^i \psi - \bar \psi \gamma^i\partial_i \psi\right) - \frac 14 \bar \psi  w_\mu^{\phantom \mu ab} \{\gamma_{ab},\gamma^\mu\} \psi \right)\,. \label{symham}
\end{equation}

Let us characterize the term $\Theta \equiv - \frac 14 \bar \psi  w_\mu^{\phantom \mu ab} \{\gamma_{ab},\gamma^\mu\}$, which can be regarded as the obstruction to write the lattice Hamiltonian simply as
\begin{equation}
H=\frac i2 \sum J_{mn}\left(\psi^\dagger_{m+1,n}\sigma_x\psi_{m,n} +\psi^\dagger_{m,n+1}\sigma_y\psi_{m,n}\right) + h.c.\,,
\end{equation}
as found for the special cases discussed in the previous sections. Using the relation between the spin-connection and the driebein and the anticommutator $\{ \gamma_{ab},\gamma^\mu\}= 2 e^{\mu c} \gamma_{abc}$ --in 2+1 dimensions it reduces to $\{ \gamma_{ab},\gamma^\mu\}= - \frac 13 e^{\mu c} \epsilon_{abc}$, it follows that
\begin{equation}
 \Theta=\frac 1{12} e^{\mu a} e^{\nu b} \partial_\mu  e_\nu^c \,\epsilon_{abc}=\frac 1{12}e^{i a} e^{\nu b} \partial_i  e_\nu^c \,\epsilon_{abc}\,.
\end{equation}
Hence, we conclude that $\Theta$ is identically zero when the metric is diagonal, as for the Rindler metric (\ref{mrindler}) and the spatially deformed metric (\ref{spam}). In order to be as general as possible, we observe that any 2+1 spacetime admitting  a timelike Killing vector can be always reduced, at least locally, to the form
\begin{equation}
 ds^2=- f^2(x,y) (dt + \lambda(x,y))^2 + f^{-2}(x,y)e^{2\Phi(x,y)}(dx^2+dy^2)\,,\label{genmetric}
\end{equation}
where $\lambda(x,y)=\lambda_x(x,y) dx + \lambda_y(x,y) dy$ is one-form independent of time.

It follows that:
$$e^0=f (dt+\lambda)\,, \ \ \ e^i= f^{-1} e^\Phi dx_i\,, \ \ \ \ x_i=x,y,\ \ \ i=1,2\,, $$
and that the inverse driebein are
$$
e_0= f^{-1} \partial_t\,, \ \ \ e_1=- f e^{-\Phi} \lambda_x \partial_t + f e^{-\Phi} \partial_x\,, \ \ \ e_2=- f e^{-\Phi} \lambda_y \partial_t + f e^{-\Phi} \partial_y
$$
The spin-connection can be chosen to be:
\begin{eqnarray}
\fl&&  w^{01}=fe^{-\Phi} \partial_x f dt +\frac{f e^{-\Phi}}2 \left(\partial_x(f\lambda_y)-\partial_y(f\lambda_x)\right) dy \cr
\fl&&  w^{02}=fe^{-\Phi} \partial_y f dt -\frac{f e^{-\Phi}}2 \left(\partial_x(f\lambda_y)-\partial_y(f\lambda_x)\right) dx \cr
\fl&&  w^{12}=\frac{f^{-1}e^{-\Phi}}{2}\frac{f e^{-\Phi}}2 \left(\partial_x(f\lambda_y)-\partial_y(f\lambda_x)\right) dt \cr 
\fl&& \qquad+ \left(\partial_y(\Phi - \ln f ) - \frac{f^2 e^{-\Phi}}2 \lambda_x \left(\partial_x(f\lambda_y)-\partial_y(f\lambda_x)\right)\right) dx  \cr
\fl       & &\qquad
           -\left(\partial_x(\Phi - \ln f )+ \frac{f^2 e^{-\Phi}}2 \lambda_y \left(\partial_x(f\lambda_y)-\partial_y(f\lambda_x)\right)\right) dy\,.
\end{eqnarray}
 This implies the following relation for the coefficients:
$$w_y^{\phantom y 01}=-w_x^{\phantom x 02} =f^{-2} e^\Phi w_t^{\phantom t 12}\,, $$
and that $w_y^{\phantom y 12}$ can be obtained from $w_x^{\phantom x 12}$ by exchanging $x$ with $y$ and 1 with 2 (which amounts for the minus sign), accordingly to the symmetry of the metric (\ref{genmetric}).

We are ready to compute $\Theta$. After some algebra we get
\begin{equation}
 \Theta = \frac{f^3 e^{-2\Phi}}{12}\left( \partial_x \lambda_y - \partial_y\lambda_x\right)\,.
\end{equation}
The above expression means that the spin-connection will cancel in the symmetric form of the Hamiltonian if and only the metric is static, i.e. the off-diagonal terms due to $\lambda$ can be reabsorbed by a change of coordinates. To prove this we note that the condition $\Theta=0$ implies that the form $\lambda$ is exact, i.e. there exists a function $F=F(x,y)$ such that $dF= \lambda$. Indeed, this is the case as as can be seen by redefining the time coordinate as $T=t+F$, as the final metric is diagonal.

With the choice of the metric (\ref{genmetric}), the overall hopping rate $J_{mn}$ is
$$J_{mn}=e^{\Phi(x_m,y_n)}.$$

\section{The density of states}

This section is devoted to deriving the relation between the density of states as a function of the energy $\rho(E)$ and the propagator in a spacetime of dimensions $(1,d)$. In order to do so we will follow a constructive procedure. By definition we have
\begin{equation}
 \int {\rm d} E \,\rho(E)= \mbox{ \# of eigenstates of the Hamiltonian $H$}\,.
\end{equation}
We will express the number of states as the number of poles in the propagator when averaged over the set of eigenstates of $H$, $\{\ket{n}\}$, which we demand be complete and normalizable (as is the case for any sound Hamiltonian operator). Furthermore, we assume that the position operator eigenstates $\ket{\bf r}$ can be completed to give an orthonormal basis $\ket{\bf r}\ket{i}$ where $i$ encodes all internal degrees of freedom, such as the spin.  Using the theorem of residues and integrating on a rectangular contour around the real axis of width $2 \epsilon>0$ , we can write:
\begin{eqnarray}
\fl\mbox{\# e.s. of $H$}&=& \frac 1{-2 \pi i} \int {\rm d} E \sum_{n} \left(\frac 1{E+i\epsilon-E_n}-\frac 1{E-i\epsilon-E_n} \right)  \cr
\fl&=& -\frac 1\pi \mbox{Im }  \int {\rm d} E \sum_{n} \bra{n}\frac 1{E-H+i\epsilon} \ket{n}\cr
\fl&=& -\frac 1\pi \mbox{Im }  \int {\rm d} E \int {{\rm d}}^d r {{\rm d}}^d r'\sum_{n,i,i'} \bra{n}\ket{\bf r}\ket i \bra{\bf r}\bra i \frac 1{E-H+i\epsilon} \ket{\bf r'} \ket{i'}\bra{\bf r'}\bra{i'}\ket{n}\cr
\fl&=& -\frac 1\pi \mbox{Im }  \int {\rm d} E \int {{\rm d}}^d r \bra{\bf r}\bra i \frac 1{E-H+i\epsilon} \ket{\bf r} \ket{i}\,.
\end{eqnarray}
Form the above equation we conclude that
\begin{equation}
\rho(E)= \frac 1\pi \mbox{Im }\int {{\rm d}}^d r \bra{\bf r}\bra i \frac 1{H-E-i\epsilon} \ket{\bf r} \ket{i}\,,
\end{equation}
that is the equation 34  given in \cite{Sitenko2007241}. To warm up we compute the density of state in the free case. As the Hamiltonian is diagonal in momentum space, it is convenient to write $\rho(E)$ in this basis:
\begin{eqnarray}
\fl\rho(E)&=&\frac 1\pi \mbox{Im }  \int {\rm d}_2\Omega_p \left(\bra{\bf p , +}  \frac 1{E-H+i\epsilon} \ket{\bf p ,+} + \bra{\bf p , -}  \frac 1{E-H+i\epsilon} \ket{\bf p ,-}\right)\cr
\fl&=& \frac 1\pi \mbox{Im }  \int {\rm d}^2 r  \int {\rm d}_2 \Omega_p \left(\bra{ {\bf p} , +} \ket{ {\bf r} } \frac{| {\bf p} | + E}{| {\bf p} |^2 - E^2 - i \mbox{sign}(E) \epsilon}\bra{ {\bf r} } \ket{ {\bf p} ,+} \right.\cr
\fl && \qquad \qquad \qquad \qquad  \left.+ \bra{ {\bf p} , -} \ket{ {\bf r} } \frac{-| {\bf p} | + E}{| {\bf p} |^2 - E^2 - i \mbox{sign}(E) \epsilon}\bra{ {\bf r} } \ket{ {\bf p} , -}\right)\cr
\fl&=& \frac 1\pi \mbox{Im }  \int {\rm d}^2 r \int \frac{{\rm d}^2 p}{(2 \pi)^2} \frac{2E}{| {\bf p} |^2 - E^2 - i \mbox{sign}(E) \epsilon}\,\label{rhofree}\,,
\end{eqnarray}
where ${\rm d}_2\Omega_p= \frac{{\rm d}^2 p}{(2\pi)^2 2 | \bf p |}$ is the Lorentz invariant measure and the normalization of the momentum states $ \ket{ {\bf p} ,\pm}$ is fixed accordingly.
By taking to be $S=\int {\rm d}^2 r$ the volume of the system and by going to polar coordinates one finds
\begin{equation}
 \rho(E)= \frac S{\pi}  \mbox{Im}\int_0^{+\infty} \frac{{\rm d} p}{2\pi}      \frac{2pE}{ p^2 - E^2 - i \mbox{sign}(E) \epsilon}\,.
 \end{equation}
 The last integral can be solved in many ways, for example by changing variable to $z=p^2$ and regularizing the integral with cut-off
\newpage
\begin{eqnarray}
\fl \qquad \int_0^{\Lambda^2} \frac{{\rm d} z}{2\pi} \frac{E}{z-E^2-i \mbox{sign}(E) \epsilon}&=&\frac E{2\pi} \left(\log(\frac{\Lambda^2-E^2}{E^2})+ \log\left(\frac{1-i \mbox{sign}(E) \epsilon}{1+i \mbox{sign}(E) \epsilon}\right)\right)\cr
&=&\frac E{2\pi}\left( \log(\frac{\Lambda^2-E^2}{E^2})+ 2 \pi i (\Theta(E)-\Theta(-E))\right)\,.
\end{eqnarray}
As the imaginary part is independent of the cut-off, the final result, as expected, is
\begin{equation}
\rho(E)= S \frac{|E|}\pi\,,
\end{equation}
which is in agreement with the result quoted in \cite{Sitenko2007241}.

It is worth-while to note that $\rho(E)$ can be written in terms of the Feynman propagator. This fact can be derived in a more general setting. By definition, $\frac 1{E-H+i\epsilon}$ is the Fourier transformation in time of the retarded propagator defined by the equation
\begin{equation}
 (i\partial_t -H){\cal G}_+(t-t')= \delta(t-t')\otimes \mathbb{I}_{\mbox{spin}}\,\label{retpro}
\end{equation}
with boundary conditions ${\cal G}_+(t-t')=0$ for $t<t'$. On the right-hand side of (\ref{retpro}), the identity in spin-space has been written explicitly to remind the reader that ${\cal G}_+(t-t')$  in general acts as a matrix on the internal degrees of freedom. After multiplying  on the left by $\gamma^t$   and on the right by $\gamma_t$ and taking the expectation value with eigenstates of the position operator the above equation gives
\begin{equation}
 -i \gamma^\mu \partial_\mu {\cal G}_+(t-t', \bf r , {\bf r'}  )\gamma_t = \delta(t-t')\delta^d({\bf r} - {\bf  r'})\otimes \mathbb{I}_{\mbox{spin}}\,,
\end{equation}
where we use $\gamma^t\gamma_t=1$.  Hence, we conclude that ${\cal G}_+(t-t', \bf r , {\bf r'}  )\gamma_t$ is related to the Feynman propagator as it solves the same equation. The precise relation can be derived by taking the boundary conditions into account. This can be explicitly checked by Fourier-transforming to momentum space. Indeed,
\begin{eqnarray}
\fl \qquad \frac 1{E-H+i\epsilon}\gamma_t&=&\frac{E-\gamma_t\gamma^i p_i}{E^2-  |{ \bf p }|^2+i\mbox{sign}(E)\epsilon} \gamma_t= \frac{E \gamma_t - \gamma^i p_i}{E^2-  |{ \bf p }|^2+i\mbox{sign}(E)\epsilon}\cr
\fl \qquad &=&\frac {1}{\gamma^\mu p_\mu-i\mbox{sign}(E)\epsilon}.
\end{eqnarray}
The above relation between the retarded propagator and Feynman propagator can be extended using perturbation theory to the interactive second quantized formalism. To conclude, let us remark that it is better to use the local definition of density (\ref{localdensity}) because it is easy to make it generally covariant in order to apply it in a curved gravitational background. In this way, the generalize notion of inner product is properly taken into account due to (\ref{curvedpro}).

\section{The computation of $\Gamma$}\label{The computation of Gamma}

In order to compute $\mbox{ Im}\, \Gamma(w, {\bf p} )$ we note that the integral of (\ref{Gamma}) can be split in
\begin{equation}
\fl  \Gamma(w, {\bf p})= 4 w \int \frac{{\rm d}^2 k}{(2\pi)^2} \left(\frac  1{{\bf k}^2 -w^2 -i\epsilon}+ \frac{w^2-\frac 34 {\bf p}^2 + {\bf p} \cdot {\bf k}}{({\bf k}^2 -w^2 -i\epsilon)(({\bf k} - {\bf p} )^2 -w^2- i\epsilon) }\right).
\end{equation}
The first integral has been computed above for the free case and its imaginary part gives
\begin{equation}
 \mbox{ Im}\, \int \frac{{\rm d}^2 k}{(2\pi)^2} \frac  {4 w}{{\bf k}^2 -w^2 -i\epsilon}= 2 w \,.
\end{equation}
The second integral is convergent and can be computed using
Feynman parameter. After some algebra one gets
\begin{eqnarray}
\fl \qquad 4 w \mbox{ Im} \int \frac{{\rm d}^2 k}{(2\pi)^2} \frac{w^2-\frac 34 {\bf p}^2 + {\bf p} \cdot {\bf k}}{({\bf k}^2 -w^2 -i\epsilon)(({\bf k} - {\bf p} )^2 -w^2- i\epsilon) }=\cr
\fl \qquad \qquad = -\frac{w}{\pi p} \mbox{Im}\left(\sqrt{4w^2+i\epsilon -p^2}\,\arctan\left[\frac{p}{\sqrt{4w^2+i\epsilon -p^2}]}\right]\right)\cr
\fl \qquad \qquad = -\frac{w}{\pi p}\sqrt{|4 w^2 -p^2|}\left(\Theta(4w^2-p^2)\mbox{ Im}\arctan\left[\frac{p}{\sqrt{4w^2+i\epsilon -p^2}]}\right] + \right.\cr
\left.
\fl \qquad \qquad \qquad \qquad \qquad - \Theta(-4w^2+p^2) \mbox{ Re}\arctan\left[\frac{p}{\sqrt{4w^2+i\epsilon -p^2}]}\right]\right),
\end{eqnarray}
where $\sqrt{4w^2+i\epsilon -p^2}=
\sqrt{|4 w^2 -p^2|}\left(\Theta(4w^2-p^2) - i \Theta(-4w^2+p^2)\right)$ and $p$ is the modulus of ${\bf p}$, $p=|{\bf p}|$.

\end{document}